\documentclass[twoside,a4paper,11pt]{sca}

\def\ga{\mathrel{\mathchoice {\vcenter{\offinterlineskip\halign{\hfil
$\displaystyle##$\hfil\cr>\cr\sim\cr}}}
{\vcenter{\offinterlineskip\halign{\hfil$\textstyle##$\hfil\cr
>\cr\sim\cr}}}
{\vcenter{\offinterlineskip\halign{\hfil$\scriptstyle##$\hfil\cr
>\cr\sim\cr}}}
{\vcenter{\offinterlineskip\halign{\hfil$\scriptscriptstyle##$\hfil\cr
>\cr\sim\cr}}}}}
\usepackage{graphicx}
\usepackage{hyperref}
\usepackage{movie15}
\usepackage[authoryear]{natbib}
\begin{document}
\pagenumbering{arabic}
\pagestyle{myheadings}
\thispagestyle{empty}
%%%%{\flushright\includegraphics[width=\textwidth,bb=58 650 590 680]{stamp.pdf}}
{\flushright\includegraphics[width=\textwidth,viewport=90 650 520 700]{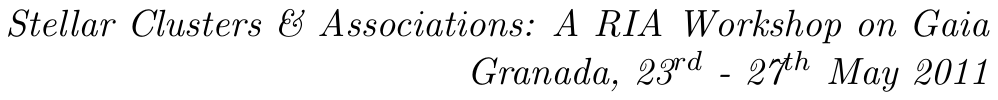}}
\vspace*{0.2cm}
\begin{flushleft}
{\bf {\LARGE
%
%%% TITLE of the paper. 
%%% TITLE of the paper. 
High-mass stars in clusters and associations
%
% Do not delete next few lines
}\\
\vspace*{1cm}
%
%%% Include here the LIST OF AUTHORS.
%%% Include here the LIST OF AUTHORS.
%%% Note that the last author has to be preceeded by an AND.
Ignacio Negueruela$^{1}$,
%
% Do not delete next few lines
}\\
\vspace*{0.5cm}
%
%%% AFFILIATIONS LIST.
%%% and the AFFILIATIONS LIST. Note that one affiliation per line.
%%% Add as many affiliations as necessary. 
$^{1}$
Departamento de F\'{i}sica, Ingenier\'{i}a de Sistemas y
  Teor\'{i}a de la Se\~{n}al, Universidad de Alicante, Apdo. 99, E03080
  Alicante, Spain\\
%
% Do not delete next few lines
\end{flushleft}
%
% Headings
\markboth{
%%% Type the SHORT version of the paper title.
%%% Type the SHORT version of the paper title.
Massive stars in clusters
}{ % Do not delete
%
%%%  First Author \& Second Author   OR   First-author et al. 
%%%  First Author \& Second Author   OR   First-author et al. if the author list 
%%% contains three or more authors.
I. Negueruela
% 
% Do not delete next few lines
}
\thispagestyle{empty}
\vspace*{0.4cm}
\begin{minipage}[l]{0.09\textwidth}
\ 
\end{minipage}
\begin{minipage}[r]{0.9\textwidth}
\vspace{1cm}
\section*{Abstract}{\small
%
% ABSTRACT ABSTRACT ABSTRACT
% ABSTRACT ABSTRACT ABSTRACT
%%% Type the ABSTRACT of your paper
High-mass stars are major players in the chemical and dynamical evolution of galaxies. Open clusters and associations represent the natural laboratories to study their evolution. In this review, I will present a personal selection of current research topics that highlight the use of open clusters to constrain different properties of high-mass stars, such as the possible existence of an upper limit for the mass of a star, the evolutionary stage of blue supergiants or the characterisation of supernova progenitors.
%
% Do not delete next few lines
\normalsize}
\end{minipage}
%
%
%%% BODY of the paper
%%% BODY of the paper
%
\section{Introduction \label{intro}}
The study of high-mass stars (also, though not quite correctly, known as massive stars) is intimately linked to, and difficult to disentangle from, the study of young open clusters. This is not only because most high-mass stars are found within young open clusters and associations, but also because clusters are the natural laboratories for investigating the evolution of high-mass stars. At the most basic level, determining the fundamental parameters of a massive star requires knowledge of its distance \citep[e.g.,][]{her92}. Since very few high-mass stars have accurate parallaxes \citep{maiz08}, membership in a cluster or association is required for calibration. In view of such a close connection, this short review can only cover a very small fraction of the many topics subject to active study that I could have considered for inclusion. I will concentrate on a few issues of current research that highlight the importance of open clusters as laboratories for understanding massive star evolution. The selection of these topics is undoubtedly biased by my personal preferences, but is representative of the areas currently generating the highest interest among researchers in the field.

\subsection{Definition of high-mass stars}
There are several possible definitions for high-mass stars, all of them indirect. We can define high-mass stars as those initiating C burning non-explosively in their cores. Modern theoretical models including overshooting indicate that this will happen for $M_{*}\ga 8\,M_{\odot}$ \citep[e.g.,][]{et04}. We can also define high-mass stars as those ending up their lives in supernova explosions. These two definitions are almost identical, though the latter implies a slightly higher mass \citep[see, e.g.,][for a description of the physics involved]{poel08}. The limit for a star to produce a supernova explosion has recently been set from observations of supernova progenitors at $\ga 8.5^{+1}_{-1.5}\,M_{\odot}$ \citep[and see also Sect.~\ref{end}]{smarttal09}. A third possible definition of high-mass stars is those stars with self-initiating radiation-driven winds \citep{kp00}. Radiative winds become detectable for main-sequence stars with spectral type earlier than B2\,V, which again roughly corresponds to $M_{*}\ga 8\,M_{\odot}$. 

Observationally, high-mass stars comprise the OB stars (approximately, O2--B2\,V, O2-B5\,III and O2-B9\,I; \citealt{reed,wf90}) and some later type supergiants (the most luminous supergiants of spectral types A, F, G, and K, and the M-type supergiants). For an observational review, see \citet{massey03}.

\subsection{Formation of high-mass stars}

The formation of high-mass stars is a major research topic in modern astrophysics, and will not be discussed here. See \citet{zy07} for a recent review. High-mass stars may have an effect on the formation of low-mass stars, and thus on the Initial Mass Function (IMF). In addition, high-mass stars may play a role in triggering further star formation. This issue has been controversial since the emergence of the classical theory \citep{el77}, because a causal relationship is difficult to assess, even if indications of sequential formation are widespread and strong indirect evidence for triggering has been found \citep[e.g.,][]{walborn02a,zavagno05}. I will not discuss star formation here, but simply note that sequential star formation within a cluster or association may give rise to populations with different ages, which might be difficult to disentangle, meaning that accurate ageing is not always possible for young open clusters.

\section{The most massive stars}

The possible existence of an upper limit to the IMF has been a hotly debated issue in recent times. Several stars believed to be extremely luminous, and thus extremely massive, have been resolved into two or more components \citep[e.g.,][]{maiz07}. Based on an analysis of the stellar population in the compact young Arches open cluster, \citet{figer05} concluded that there was clear evidence for an upper limit to stellar masses around $150\,M_{\odot}$. This result is dependent on several factors, such as the extinction law to the Galactic Centre, but has also been reproduced for the Large Magellanic Cloud \citep{koen06}. 

Recent modelling advances have revealed that hydrogen-rich Wolf-Rayet (WR) stars of the nitrogen sequence (WN) are really main-sequence objects with heavy mass loss. These are candidates to be the most massive stars in their hydrogen core burning phase \citep{langer94,dekoter97,cd98}. The fact that they are commonly found heavily packed at the core of the densest young open clusters supports this view. Searches for very massive stars have thus concentrated on possible binaries amongst this population, as close binaries may provide accurate masses via dynamical studies.

One of the best examples is found in the obscured starburst cluster Westerlund~2. Located in the Sagittarius Arm (likely at $\sim8$~kpc, though distances as short as 3~kpc have been given), this cluster contains at least 
14 early O or WR stars \citep{rauw07}. Its total mass is also open to discussion, but may approach $M_{{\rm cl}}=10^4\,M_{\odot}$ \citep{ascenso07}, a result supported by the recent discovery of two very massive stars likely ejected from the cluster core \citep{rl11}. One of the most luminous members of Westerlund~2, the WN6h star WR20a, was identified as a massive binary by \citet{rauw04} and as an eclipsing binary by \citet{bonanos04}. Combination of the radial velocity curve and the lightcurve provides an accurate solution with $P_{{\rm orb}} = 3.69\pm0.01$~d. The components have masses $M_{1} =83\pm5\,M_{\odot}$ and $M_{2} =82\pm5\,M_{\odot}$. They have the same spectral type and the same mass, which is the highest stellar mass measured with high accuracy.  

Even more massive than Westerlund~2, the cluster at the core of the  giant H\,{\sc ii} region NGC 3603, also located in the Sagittarius arm (at $\sim7$~kpc), is extraordinarily compact. It contains 
$\approx 35$ early O or WR stars \citep{moffat94} and likely has a mass $M_{{\rm cl}}=1.5\times10^4\,M_{\odot}$ \citep[and references therein]{rochau10}. Its central concentration contains 3 WN6ha stars, the brightest of which is an eclipsing binary with $P_{{\rm orb}} = 3.7724$~d. The orbit has been solved using $K$-band spectroscopy \citep{schnurr08}. The radial velocity curve of the secondary has large uncertainties and therefore the masses are not very tightly constrained. The solution implies  $M_{1} =116\pm31\,M_{\odot}$ and $M_{2} =89\pm16\,M_{\odot}$. Component 1 is thus the most massive star with a dynamically determined mass \citep{schnurr08}. 

Outside the Milky Way, the 30 Dor complex in the Large Magellanic Cloud is likely the most massive starburst in the Local Group. Its nuclear cluster, Radcliffe 136, is 2.7~pc across and contains hundreds of OB stars. Its mass has been estimated to be, at least, $M_{{\rm cl}}=1\times10^5\,M_{\odot}$ \citep{andersen09}. Using updated stellar models, \citet{crowther10} found that the dynamical masses of the WN6ha stars in NGC~3603 could be well reproduced by evolutionary tracks. If these models are then extrapolated to the even brighter WN5h stars at the core of 30~Dor, they indicate enormous masses. Three of the WN5h stars are fitted with models implying current masses $\geq 150\,M_{\odot}$, with Star~a1 having $265^{+80}_{-35}\,M_{\odot}$ \citep{crowther10}. This would be the most massive star known, well above the proposed upper mass limit. Unfortunately, none of the stars at the core of 30~Dor appears to be a spectroscopic binary to confirm this determination with a dynamical measurement. 

\section{Stellar evolution}

High-mass stars play a decisive role in driving the chemical evolution of galaxies. Heavy mass loss through all their life stages and in the final supernova explosion provides an important fraction of the heavy elements in the interstellar medium. In order to understand the chemical enrichment of the medium, we need to constrain how and when mass is lost by high-mass stars, and this means being able to map observed phases on to theoretical evolutionary tracks.
 
Open clusters and associations allow us to explore the evolutionary context of massive stars in different evolutionary stages \citep[e.g.,][]{walborn10}. This process starts with the youngest stellar systems. For example, the Carina nebula provides many of the MK standards for the earliest O subtypes \citep{walborn02}. The fact that O3--4 stars are still on the main sequence indicates that the complex is very young, while this very youth constrains its only evolved star, $\eta$~Car, to be extremely massive.

\begin{figure}
\center
\includegraphics[width=12cm,clip=true]{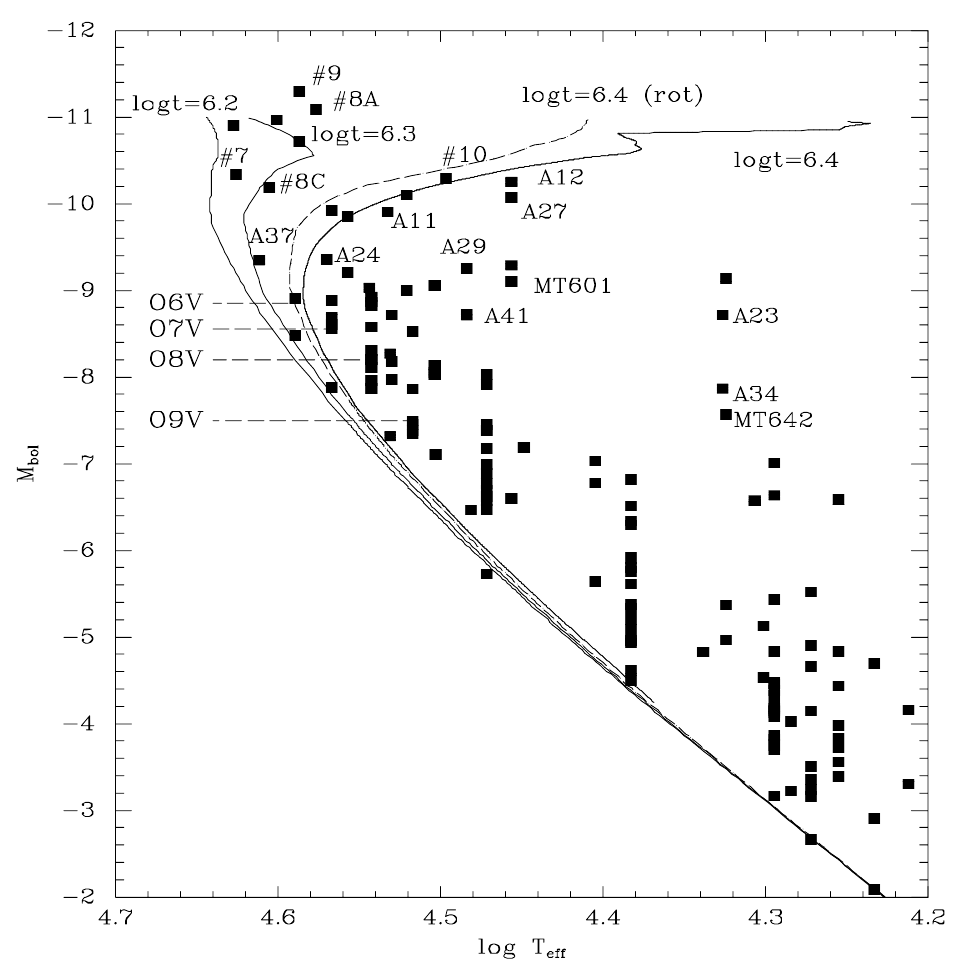}
\caption{\label{hr} Semi-empirical HR diagram for Cyg~OB2, after \citet{neg08}.  Note that, in the original paper, the $(V-K)_{0}$ term was added twice, because of a mistake with the spreadsheet. When this is taken into account, the distance modulus $DM=10.8$ proposed by \citet{han03} is no longer favoured. A better fit is now
obtained with the classical value $DM=11.3$, as displayed.
Continuous lines are non-rotating isochrones for $\log
   t=6.2$, $6.3$ and $6.4$ from \citet{schaller}. The dashed line is the
   $\log t=6.4$ isochrone in the  high-rotation models
   \citep{mm03}.
The main-sequence turn-off at O6\,V seems to be followed by a number of evolved stars tracing the $\log
t=6.4$ isochrone. A number of O3--5 supergiants, though, fall well to the left of the isochrone.}
\end{figure}

 At a slightly older age, Cyg~OB2 represents an important laboratory for the study of O-type stars. Considered a very promising target since its discovery \citep{jm54}, it could only be studied in detail after the advent of CCD detectors because of the high obscuration, in spite of its relative low distance (see Fig.~\ref{hr}). \citet{mt91} identified $\sim60$ stars more massive than $15M_{\odot}$, finding that Cyg~OB2 is very compact for an OB association and seems to 
occupy a more or less unique position somewhat 
intermediate between an open cluster and a normal OB association
\citep[cf.][]{kno00}. The central region contains two cluster-like
stellar concentrations \citep{bica03}. They form an elongated figure
$\sim4^{\prime}\times10^{\prime}$ ($\sim2\times5$~pc at the distance
of Cyg~OB2) and contain 18 O-type stars (assuming that all systems are single; given the observed multiplicity -- see below --, the actual number is likely to be a factor of two higher).

After analysis of the 2MASS data in this region suggested a much
richer population of early type stars \citep{kno00}, several recent studies \citep[e.g.,][]{com02,han03} have shown that the association contains $>50$ (and likely $\sim70$--80) O-type stars \citep[e.g.,][]{kiminki07,neg08}. The main sequence is very well defined, showing a clear turn-off at O6\,V. A number of giants and supergiants follow the 2.5~Myr isochrone from this turnoff. However, there is also a population of O3--O5 supergiants, which seems incompatible with this age \citep[see][and Fig~\ref{hr}]{neg08}. With the data available, it is impossible to rule out the possibility that these stars represent a younger population. However, the similarity of the HR diagram for Cyg~OB2 to theoretical plots in which very fast rotation generates blue stragglers \citep{demink} is striking.

The presence of blue supergiants with spectral types too early for the main-sequence turnoff is very frequent in young open clusters, but these objects have not generally been classified as blue stragglers, even if they are located to the left of the cluster isochrone, the classical definition of a blue straggler. A paradigmatic case is h~Persei (NGC~869), one of the most well-studied clusters in the Northern sky, and a rather massive cluster -- recent mass estimates indicate 
%$\sim7\,000\,M_{\odot}$ for cluster plus halo \citep{sles02} and
 $\sim4\,500\,M_{\odot}$ in the core region \citep{currie10}. Most recent works seem to converge on an age $\sim14$~Myr, in good agreement with an apparent main-sequence turn-off at B1\,V. The luminous supergiants in the cluster, however, have spectral types too early for this age \citep{marco01}, at 
B2\,Ia (HD~14143) and B3\,Ia (HD~14134), and evolutionary masses approaching $40\,M_{\odot}$ \citep{mcerlean,crowther06a}. A similar situation is found in many other clusters \citep{marco07}. In most cases, there are no observational reasons to suggest that these supergiants are members of a younger (second-generation) population. This clearly shows that we are still very far from understanding exactly which evolutionary phase blue supergiants represent.

Because of the huge size of its stellar population, the young cluster Westerlund~1 (Wd~1) is currently our best laboratory for studying the evolution of high-mass stars. Located at $\sim5$~kpc, its mass exceeds $5\times10^{4}\:M_{\odot}$ \citep{gennaro11} and may approach $10^{5}\:M_{\odot}$. With more than twenty WR stars \citep{crowther06b} and several dozen OB supergiants \citep{neg10}, Wd~1 provides stringent tests on current theoretical models. The population observed is consistent with a single burst of star formation. There may be some mild blue stragglers, but the blue supergiants in the cluster (including some very luminous late-B hypergiants with $M_{V}\sim-9$) fit rather well the theoretical isochrones for ages between 5 and 6~Myr \citep{neg10}. Moreover, the dynamical determination of the mass for the two components of the eclipsing binary Wd1-W13 shows good agreement with the expected theoretical masses, with the B1\,I component having a mass $M_{*}=35\pm5\,M_{\odot}$ \citep{ritchie10}. This good agreement is comforting in view of the difficulties in understanding blue supergiants in other clusters and provides strong support for the basic assumptions underlying current evolutionary tracks \citep[e.g.,][]{mm00}.

\section{Binarity}

The large number of high-mass stars present in Wd~1 offers an ideal opportunity to study their binary fraction. The population of WR stars offers strong indirect evidence for a very high binary fraction \citep{crowther06b}. Targeted observations of the blue supergiant population revealed that at least 40\% of the stars observed are binaries \citep{ritchie09}. Observations of a larger sample, including stars closer to the main sequence, are in progress.

Searches for binaries in Cyg~OB2 are, to date, more complete. A radial velocity survey carried out over several years \citep{kiminki07, kobu11} should be able to reveal, except for very unfavourable inclinations, all massive companions and a substantial fraction of low-mass companions. The survey has so far detected 20 spectroscopic binaries, while 
20 other objects show radial velocity variations, and might be binaries.
Comparison to theoretical expectations suggests a very high binary fraction, perhaps approaching unity. The masses of the companions seem to follow a flat distribution \citep{kiminki09, kobu11}.

The binary fraction and companion mass distribution do not seem to be universal, though. Variability in both observables is high among the few clusters with dedicated studies. In NGC~6231, the complete sample of O-type stars (16 objects) presents $f_{{\rm bin}} > 0.63$, with companions having consistently $M_{*} > 5\:M_{\odot}$ \citep{sana08}. In contrast, the smaller cluster NGC~2244, with 6 O-type stars, has a much smaller binary fraction, $f_{{\rm bin}} > 0.17$, while in IC~1805 (10 O-type stars) $f_{{\rm bin}} > 0.20$, with the possibility of massive companions almost ruled out \citep{mahy09, debecker06}.

Meanwhile, spectroscopic monitoring of a large sample of O and WN stars has revealed a very high fraction of spectroscopic binaries \citep{barba10}. Interferometric observations, which sample a different range of orbital sizes, have also demonstrated a very high degree of multiplicity amongst high-mass stars. O-type stars in clusters display a much higher binary fraction than field O stars \citep{mason}. This can be naturally explained if many (most) field stars have been ejected from open clusters (in accordance with theoretical predictions presented in several contributions to these proceedings).

\section{Cool massive stars\label{end}}

After the blue supergiant phase, the evolutionary tracks of massive stars become more uncertain. An important fraction of high-mass stars (perhaps all with $M_{*}\ga25\:M_{\odot}$) become WR stars. Before reaching this hydrogen-depleted phase, the stars must lose their envelopes, but the evolutionary phase when this mass loss takes place is unclear. Wd~1 contains a high number of yellow hypergiants (with spectral types A and F), which seem to be subject to heavy pulsations \citep{clark10}. Current evolutionary tracks would predict these objects to become Luminous Blue Variables (LBV) and then reach the WR stage. Nevertheless, the cluster also contains a number of red supergiants, which are not predicted to exist at such high luminosities \citep{clark10}.

Stars less massive than $\sim 25\,M_{\odot}$ are expected to end their lives as red supergiants (RSGs). Modelling of RSGs is very complex due to a number of factors, among which we can cite their huge size, poorly determined molecular opacities and the expected heavy mass loss. Because of these difficulties, their fundamental properties are not very well known, though important advances have been obtained with new model fits to a sample of RSGs in open clusters and associations \citep{levesque05}. Spectroscopic monitoring of RSGs in clusters has allowed the detection of irregular radial velocity variations, which may be the signature of pulsation, the most likely source of mass loss \citep{mermilliod08}.

\begin{figure}
\center
\resizebox{\textwidth}{!}{
\includegraphics[]{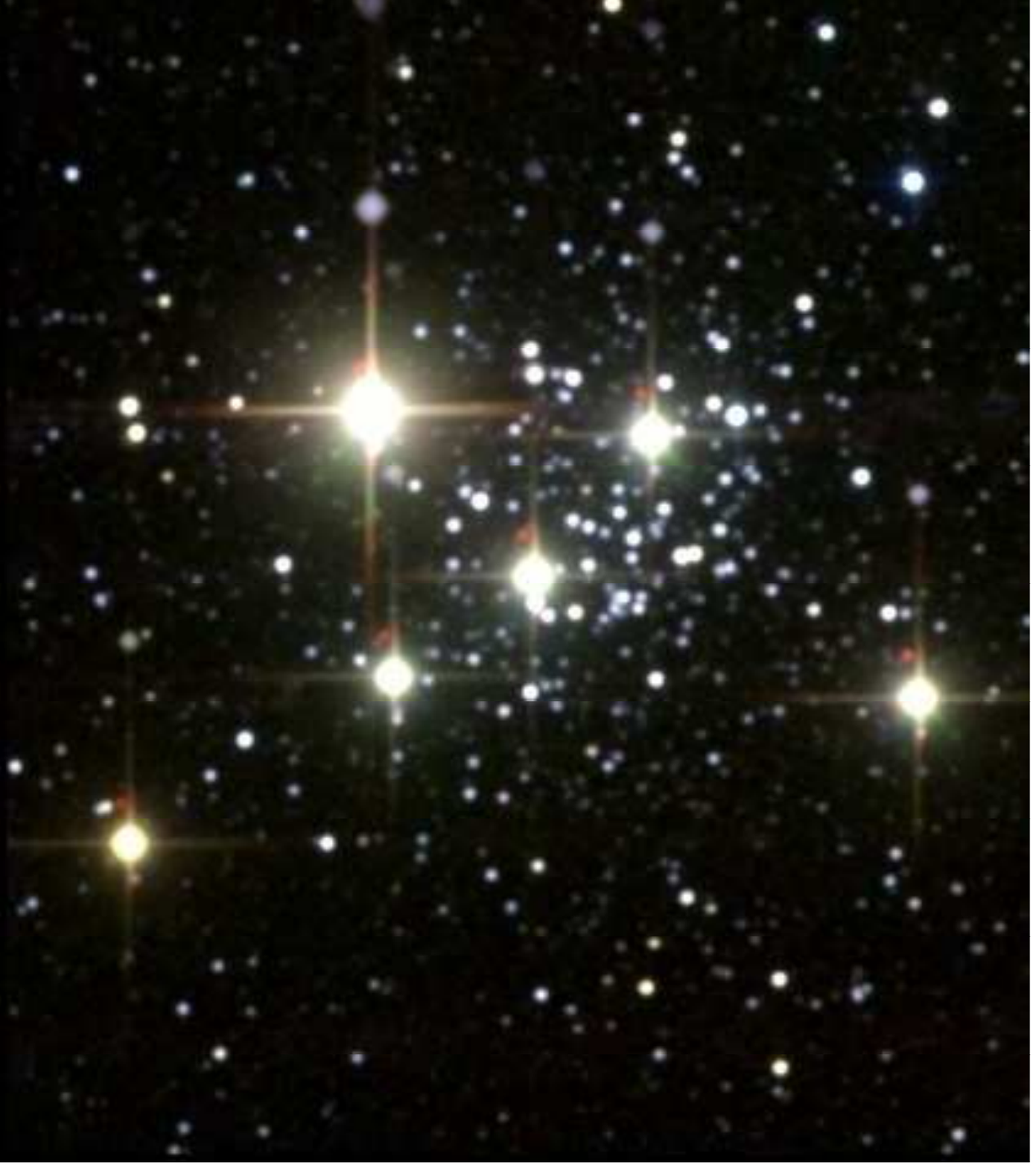} ~
\includegraphics[]{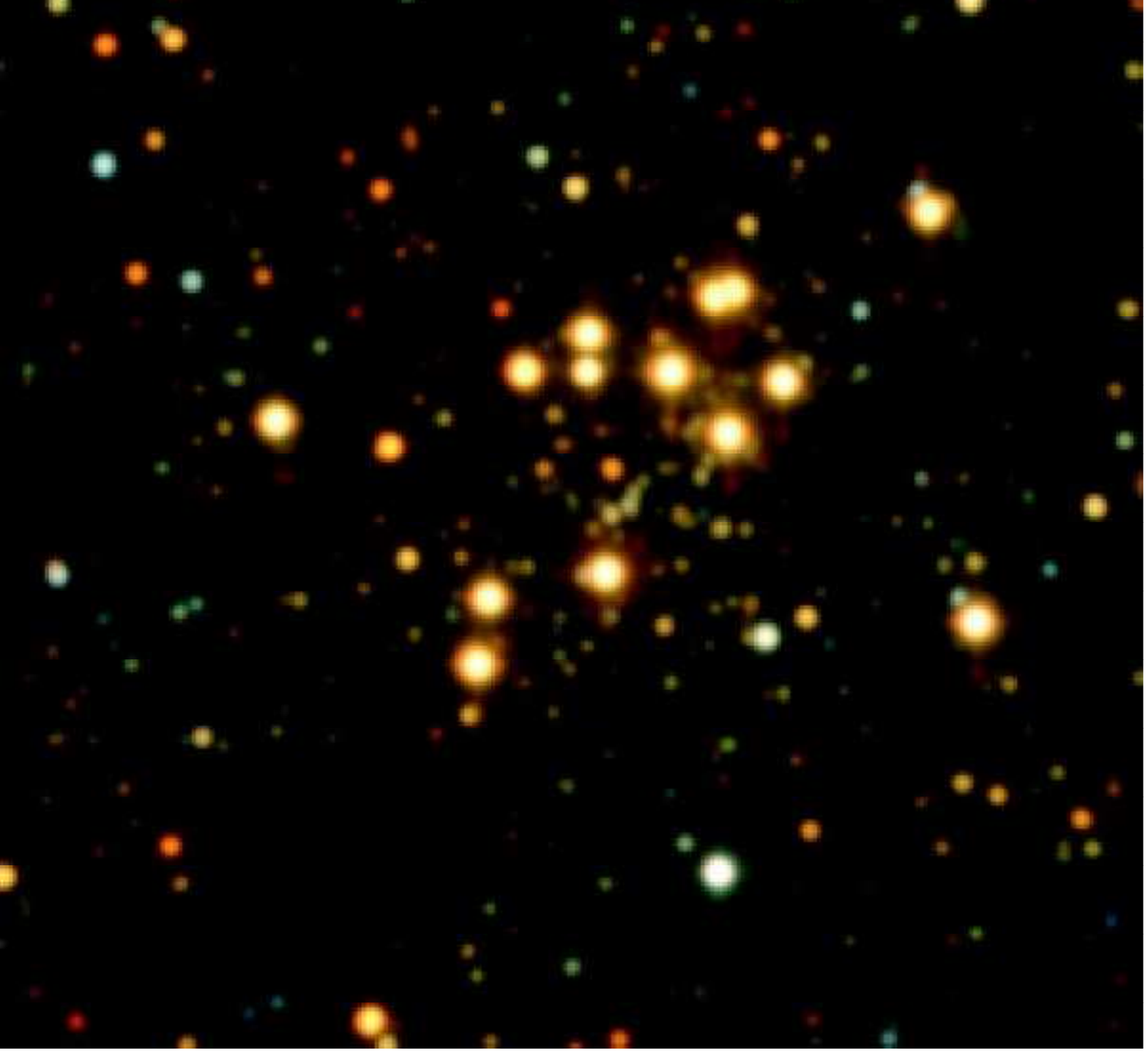} 
}
\caption{\label{ir} Left: 2MASS image of the open cluster NGC~7419. This obscured cluster in the Perseus Arm contains 5 red supergiants for no blue supergiants and presents the highest fraction of Be stars among Galactic clusters \citep{marco11}. Current theoretical models predict that this cluster should have a mass approaching $10^{4}\:M_{\odot}$ to contain 5 RSGs \citep{clark09b}, but its mass is unlikely to exceed $3\times10^{3}\:M_{\odot}$. Right: 2MASS image of RSGC1, a cluster of red supergiants at $d\sim6$~kpc, in the Inner Galaxy \citep{figer06}. This cluster has an age $\sim10$~Myr and may have $\geq3\times10^{4}\:M_{\odot}$ \citep{davies08}. Comparison of this image to the one in the left panel highlights the brightness of RSGs in the near infrared and their role as signposts of recent star formation. The population of blue stars easily seen in NGC~7419 fades out in RSGC1 due to the higher distance, much higher extinction, and confusion.
}
\end{figure}

In recent years, RSGs have shown their usefulness as signposts of massive clusters. As they are very bright in the infrared, they can be seen through heavy absorption. Several clusters rich in RSGs have been discovered towards the inner Galaxy \citep[e.g.,][]{figer06, clark09}. Determination of their chemical composition has provided extremely valuable clues to the Galactic chemical distribution \citep{davies09}. These studies are allowing the discovery of an increasing number of massive young open clusters in the Galaxy \citep[see][for a review]{pz10}. Unfortunately, the actual masses of these obscured clusters are not known, as we can only see the RSGs and not the associated main sequence (see Fig.~\ref{ir}). The most extreme case is the open cluster Stephenson~2, which contains at least 26 RSGs \citep{davies07}, and might have a mass approaching $10^{5}\:M_{\odot}$, if current predictions for the duration of the RSG phase are correct. Unfortunately, this is not certain, as the case of the open cluster NGC~7419 \citep[and Fig.~\ref{ir}]{marco11} highlights.

One important test for stellar evolution models is the ratio of blue to red supergiants in a population. This ratio is very sensitive to input physics, such as treatment of mass loss, convection and mixing processes. Unfortunately, in most Galactic clusters, the number of supergiants is so small that the ratio is not statistically significant. Studies so far have considered average values over a number of open clusters, finding an apparent dependence on Galactocentric radius. When observations of the Magellanic Clouds are also included, there is strong evidence for an increase in the blue to red supergiant ratio with increasing metallicity \citep{eggenb02}. This is exactly the opposite behaviour to what current evolutionary tracks would predict \citep{meynet11}.

However, simply counting all red and blue supergiants may provide a distorted picture, as not all ``supergiants'' sample the same population.
As an example, the open cluster NGC~6649, located at $\sim2$~kpc in the Sagittarius arm, has an age $\sim50$~Myr \citep{turner81} and contains at least three low-luminosity RSGs (with spectral types around K1\,Ib) and one supergiant Cepheid variable. According to modern evolutionary tracks \citep[e.g.,][]{marigo08}, the progenitors of these supergiants had initial masses $\sim 8\,M_{\odot}$. Stars of such low masses should never appear as blue supergiants (according to the same tracks), and so including them in counts to determine the red to blue ratio may be misleading.

Nevertheless, these low-mass RSGs are very interesting objects in themselves, as they must be the progenitors of most Type~II supernovae. This is a simple consequence of the shape of the IMF, as there must be many more stars with $M_{*}\sim8\,M_{\odot}$ than with $M_{*}\sim20\,M_{\odot}$. As Type~II supernovae are very important contributors to the dynamical and chemical evolution of galaxies, determining the lower mass for a supernova explosion becomes a fundamental issue to understand the history of the Universe. In recent years, an important observational effort has been dedicated to identifying supernova progenitors and deriving their properties \citep[see][]{smartt}. In some cases, it has been possible to identify the actual star in pre-explosion images. Sometimes, the supernova seems to have taken place within an open cluster or association and properties can be inferred from the parent population. 

Based on these observations, the minimum mass for an exploding star has been estimated at $\approx 8.5^{+1}_{-1.5}\,M_{\odot}$ \citep{smarttal09}. Efforts are also being dedicated to determining whether there is a direct relation between the initial mass of a star and the class of supernova explosion it will undergo. Of course, the final fate of a massive star depends on several factors, such as mass loss rates during the different evolutionary phases, 
initial rotational velocity and its evolution, and, perhaps most decisively, binary interaction \citep[e.g.,][]{meynet11}. In spite of this, attempts have been made to the develop ``typical'' scenarios \citep[e.g.,][]{smith11}. 

Another interesting fact is the apparent correlation between the fraction of Be stars among B-type stars and the number of RSGs. Both seem to increase in a similar way with decreasing metallicity \citep{meynet07}, suggesting a possible evolutionary connection with fast rotation. In this sense, the open cluster NGC~7419 (Fig.~\ref{ir}), which has the highest fraction of Be stars amongst Milky Way clusters, presents a blue to red supergiant ratio of 0/5 \citep{marco11}, atypical for Milky Way clusters. In contrast, NGC~663, which also has a very high Be-star fraction and is located at approximately the same Galactocentric distance, presents a ratio 5/0 for the core region (6/2 when the halo is also considered), showing that correlations are not always direct or simply that statistical fluctuations may dominate the observables even in moderately massive clusters.

\section{Outlook}

In the near future, {\it Gaia} is bound to play a decisive role in furthering our knowledge of high-mass stars. Accurate parallaxes will result in much better distances to open clusters, implying improved luminosity determinations. In addition, confirmation of membership for individual peculiar objects will give them an evolutionary context, providing strong constraints for evolutionary tracks. {\it Gaia} may also contribute strongly to the study of binarity amongst high-mass stars, as it will obtain accurate lightcurves for huge samples and may detect orbital motions in wide binary systems.

 Of course, {\it Gaia} cannot solve every problem. We will need other data and also new theoretical developments. High-resolution spectroscopy coupled with accurate stellar atmosphere models will be needed to determine stellar parameters beyond luminosity (see, e.g., Sim\'on-D\'{\i}az et al., in these proceedings). Improved evolutionary tracks will result from deeper interaction between theory and observations. Radial velocity curve solutions for spectroscopic binaries will sample the intermediate range of binary separations.

But, above all, we have to be conscious of one main limitation: {\it Gaia} will not be able to sample the inner Galaxy, where high extinction and crowding will impede the acquisition of high-quality data. Almost all the massive young clusters known will be beyond its reach. Fortunately, a full complement of new instruments will help us to study star formation in the inner Galaxy. Among them, the first generation of infrared spectrographs with high multiplexing capabilities, such as KMOS at the VLT or MIRADAS at GTC, will play a fundamental role in studying obscured clusters and massive star-forming regions. Meanwhile, new space telescopes will provide the tools to resolve clusters in the galaxies of the Local Group.
JWST may be an ideal tool for detecting RSGs in distant galaxies (and hence identifying supernova progenitors), while WSO/UV will be used to study low-extinction populations of massive hot stars. In all, with this formidable set of new missions and instruments, the next few years are likely to see important developments in this field.

%
% Do not delete the next line
\small  % Do not delete
%
%%% Comment the following line if you do not have acknowledgments.
\section*{Acknowledgments}   % Do not delete if you declare acknowledgments
%
%%% ACKNOWLEDGMENTS
%%% ACKNOWLEDGMENTS
This research is partially supported by the Spanish Ministerio de
Ciencia e Innovaci\'on (MICINN) under
grants AYA2010-21697-C05-05 and CSD2006-70. I thank Jes\'us Ma\'{\i}z Apell\'aniz and Sergio Sim\'on D\'{\i}az for comments on the manuscript.

The Two Micron All Sky Survey (2MASS) is a joint project of the University of Massachusetts and the Infrared Processing and Analysis Center/California Institute of Technology, funded by the National Aeronautics and Space Administration and the National Science Foundation.

%
% Do not delete the next few lines
%************************************************************************************%

%************************************************************************************%
%
% Do not delete the next few lines

%\bibliographystyle{aa}
%\bibliography{mnemonic,ref_user}

\begin{thebibliography}{}
\small
%
%% BIBLIOGRAPHY
%% BIBLIOGRAPHY

\bibitem[\protect\citeauthoryear{Andersen et al.}{2009}]{andersen09} Andersen, M., Zinnecker, H., Moneti, A., et al. 2009, ApJ, 707, 1347

\bibitem[\protect\citeauthoryear{Ascenso et al.}{2007}]{ascenso07} Ascenso, J., Alves, J., Beletsky, Y., \& Lago, M.T.V.T. 2007, A\&A, 466, 137

\bibitem[\protect\citeauthoryear{Barb\'a et al.}{2010}]{barba10} Barb\'a, R.H., Gamen, R., Arias, J.I., et al. 2010, RMxAC, 38, 30

\bibitem[\protect\citeauthoryear{de Becker et al.}{2006}]{debecker06} de Becker, M., Rauw, G., Manfroid, J., \& Eenens, P. 2006, A\&A, 456, 1121

\bibitem[Bica et al.(2003)]{bica03} Bica, E., Bonatto, Ch., \&  Dutra,
  C. M. 2003, A\&A, 405, 991

\bibitem[\protect\citeauthoryear{Bonanos et al.}{2004}]{bonanos04} Bonanos, A.Z., Stanek, K.Z., Udalski, A., et al. 2004, ApJ, 611, L33 

\bibitem[\protect\citeauthoryear{Clark et al.}{2009a}]{clark09} Clark, J.S., Negueruela, I., Davies, B., et al. 2009a, A\&A, 498, 109

\bibitem[Clark et al.(2009b)]{clark09b}
Clark, J. S., Davies, B., Najarro, F., et al. 2009b, A\&A, 504, 429

\bibitem[\protect\citeauthoryear{Clark et al.}{2010}]{clark10} Clark, J.S., Ritchie, B.W., \& Negueruela, I. 2010, A\&A, 514, A87

\bibitem[Comer\'on et al.(2002)]{com02}Comer\'on, F., Pasquali, A.,
  Rodighiero, G., et al. 2002, A\&A, 389, 874

\bibitem[\protect\citeauthoryear{Crowther \& Dessart}{1998}]{cd98} Crowther, P.A., \& Dessart, L. 1998, MNRAS, 296, 662

\bibitem[\protect\citeauthoryear{Crowther et al.}{2006a}]{crowther06a} Crowther, P.A., Lennon, D.J., \& Walborn, N.R. 2006a, A\&A, 446, 279

\bibitem[\protect\citeauthoryear{Crowther et al.}{2006b}]{crowther06b} Crowther, P.A., Hadfield, L.J., Clark, J.S., et al. 2006b, MNRAS, 372, 1407

\bibitem[\protect\citeauthoryear{Crowther et al.}{2010}]{crowther10}Crowther, P.A., Schnurr, O., Hirschi, R., et al. 2010, MNRAS, 408, 731

\bibitem[\protect\citeauthoryear{Currie et al.}{2010}]{currie10}Currie, T., Hernandez, J., Irwin, J., et al. 2010, ApJS, 186, 191

\bibitem[Davies et al.(2007)]{davies07}
Davies, B., Figer, D.F., Kudritzki, R.-P., et al. 2007, ApJ, 671, 781

\bibitem[Davies et al.(2008)]{davies08}
Davies, B., Figer, D.F., Law, C.J., et al. 2008, ApJ, 676, 1016 

\bibitem[\protect\citeauthoryear{Davies et al.}{2009}]{davies09} Davies, B., Origlia, L., Kudritzki, R.-P., et al. 2009, ApJ, 696, 2014

\bibitem[\protect\citeauthoryear{Eggenberger et al.}{2002}]{eggenb02} Eggenberger, P., Meynet, G., \& Maeder, A. 2002, A\&A, 386, 576

\bibitem[\protect\citeauthoryear{Eldridge \& Tout}{2004}]{et04} Eldridge, J.J., \& Tout, C.A. 2004, MNRAS, 353, 87 

\bibitem[\protect\citeauthoryear{Elmegreen \& Lada}{1977}]{el77} Elmegreen, B.G., \& Lada, C.J. 1977, ApJ, 214, 725 

\bibitem[\protect\citeauthoryear{Figer}{2005}]{figer05} Figer, D.F. 2005, Nature, 434, 192

\bibitem[\protect\citeauthoryear{Figer et al.}{2006}]{figer06} Figer, D.F., MacKenty, J.W., Robberto, M., et al. 2006, ApJ, 643, 1166

\bibitem[\protect\citeauthoryear{Gennaro et al.}{2011}]{gennaro11} 	
Gennaro, M., Brandner, W., Stolte, A., \& Henning, Th. 2011, MNRAS, 412, 2469

\bibitem[\protect\citeauthoryear{Hanson}{2003}]{han03} Hanson, M.M. 2003, ApJ 597, 957 

\bibitem[\protect\citeauthoryear{Herrero et al.}{1992}]{her92} Herrero, A., Kudritzki, R.-P., Vilchez, J.M., et al. 1992, A\&A, 261, 209 

\bibitem[Johnson \& Morgan(1954)]{jm54}Johnson, H.L., \& Morgan,
  W. W. 1954, ApJ, 119, 344 

\bibitem[\protect\citeauthoryear{Kiminki et al.}{2007}]{kiminki07} Kiminki, D.C., Kobulnicky, H. A., Kinemuchi, K., et al. 2007, ApJ, 664, 1102

\bibitem[\protect\citeauthoryear{Kiminki et al.}{2009}]{kiminki09} Kiminki, D.C., Kobulnicky, H.A., Gilbert, I., et al. 2009, AJ, 137, 4608 

\bibitem[\protect\citeauthoryear{Kn\"odlseder}{2000}]{kno00} Kn\"odlseder, J. 2000, A\&A, 360, 539 

\bibitem[\protect\citeauthoryear{Kobulnicky \& Kiminki}{2011}]{kobu11} Kobulnicky, H.A., \& Kiminki, D.C. 2011, BSRSL, 80, 616

\bibitem[\protect\citeauthoryear{Koen}{2006}]{koen06} Koen, C. 2006, MNRAS, 365, 489

\bibitem[\protect\citeauthoryear{de Koter et al.}{1997}]{dekoter97} de Koter, A., Heap, S.R., \& Hubeny, I. 1997, ApJ, 477, 792

\bibitem[\protect\citeauthoryear{Kudritzki \& Puls}{2000}]{kp00} Kudritzki, R.-P., \& Puls, J. 2000, ARA\&A, 38, 613

\bibitem[\protect\citeauthoryear{Langer et al.}{1994}]{langer94} Langer, N., Hammann, W.-R., Lennon, M., et al. 1994, A\&A, 290, 819

\bibitem[\protect\citeauthoryear{Levesque et al.}{2005}]{levesque05} Levesque, E.M., Massey, P., Olsen, K.A.G., et al. 2005, ApJ, 628, 973

\bibitem[\protect\citeauthoryear{Mahy et al.}{2009}]{mahy09} Mahy, L., Naz\'e, Y., Rauw, G., et al. 2009, A\&A, 502, 937

\bibitem[\protect\citeauthoryear{Ma\'{\i}z Apell\'aniz et al.}{2007}]{maiz07} Ma\'{\i}z Apell\'aniz, J., Walborn, N.R., Morrell, N.I., et al. 2007, ApJ, 660, 1480

\bibitem[\protect\citeauthoryear{Ma\'{\i}z Apell\'aniz et al.}{2008}]{maiz08} Ma\'{\i}z Apell\'aniz, J., Alfaro, E., \& Sota, A. 2008, {\tt arXiv:0804.2553}

\bibitem[\protect\citeauthoryear{Marco \& Bernabeu}{2001}]{marco01} Marco, A., \& Bernabeu, G. 2001, A\&A, 372, 477

\bibitem[\protect\citeauthoryear{Marco \& Negueruela}{2011}]{marco11} Marco, A., \& Negueruela, I. 2011, BSRSL, 80, 396

\bibitem[\protect\citeauthoryear{Marco et al.}{2007}]{marco07} Marco, A., Negueruela, I., \& Motch, C. 2007, ASPC, 361, 388

\bibitem[Marigo et al.(2008)]{marigo08}Marigo, P., Girardi, L.,
  Bressan, A., et al. 2008, A\&A, 482, 883

\bibitem[\protect\citeauthoryear{Mason et al.}{2009}]{mason} Mason, B.D., Hartkopf, W.I., Gies, D.R., et al. 2009, AJ, 137, 3358

\bibitem[\protect\citeauthoryear{Massey}{2003}]{massey03} Massey, P. 2003, ARA\&A, 41, 15

\bibitem[\protect\citeauthoryear{Massey \& Thompson}{1994}]{mt91} Massey, P., \& Thompson, A.B. 1991, AJ, 101, 1408

\bibitem[\protect\citeauthoryear{McErlean et al.}{1999}]{mcerlean} McErlean, N.D., Lennon, D.J., \& Dufton, P.L. 1999, A\&A, 349, 553

\bibitem[\protect\citeauthoryear{Mermilliod et al.}{2008}]{mermilliod08} Mermilliod, J.C., Mayor, M.,  \& Udry S. 2008, A\&A, 485, 303

\bibitem[Meynet \& Maeder(2000)]{mm00}
Meynet, G., \& Maeder, A. 2000, A\&A, 361, 101

\bibitem[Meynet \& Maeder(2003)]{mm03} Meynet, G., \& Maeder, A. 2003,
  A\&A, 404, 975 

\bibitem[\protect\citeauthoryear{Meynet et al.}{2007}]{meynet07} Meynet, G., Eggenberger, P., \& Maeder, A. 2007, IAUS, 241, 13

\bibitem[\protect\citeauthoryear{Meynet et al.}{2011}]{meynet11} Meynet, G., Georgy, C., Hirschi, R., et al. 2011, BSRSL, 80, 266

\bibitem[\protect\citeauthoryear{de Mink et al.}{2009}]{demink} de Mink, S.E., Pols, O.R., Langer, N., \& Izzard, R.G. 2009, A\&A, 497, 243

\bibitem[\protect\citeauthoryear{Moffat et al.}{1994}]{moffat94} Moffat, A.F.J., Drissen, L., \& Shara, M.M. 1994, ApJ, 436, 183

\bibitem[\protect\citeauthoryear{Negueruela et al.}{2008}]{neg08} Negueruela, I., Marco, A., Herrero, A., \& Clark, J.S. 2008, A\&A, 487, 575

\bibitem[\protect\citeauthoryear{Negueruela et al.}{2010}]{neg10} Negueruela, I., Clark, J.S., \& Ritchie, B.W. 2010, A\&A, 516, A78 

\bibitem[\protect\citeauthoryear{Poelarends et al.}{2008}]{poel08} Poelarends, A.J.T., Herwig, F., Langer, N., \& Heger, A. 2008, ApJ, 675, 614

\bibitem[\protect\citeauthoryear{Portegies Zwart et al.}{2010}]{pz10} Portegies Zwart, S.F., McMillan, S.L.W., \& Gieles, M. 2010, ARA\&A, 48, 431 

\bibitem[\protect\citeauthoryear{Rauw et al.}{2004}]{rauw04} Rauw, G., de Becker, M., Naz\'e, Y., et al. 2004, A\&A, 420, L9

\bibitem[\protect\citeauthoryear{Rauw et al.}{2007}]{rauw07} Rauw, G., Manfroid, J., Gosset, E., et al. 2007, A\&A, 463, 981

 \bibitem[\protect\citeauthoryear{Reed}{2009}]{reed} Reed, B.C. 2003, AJ, 125, 2531

\bibitem[\protect\citeauthoryear{Ritchie et al.}{2009}]{ritchie09} Ritchie, B.W., Clark, J.S., Negueruela, I., \& Crowther P.A. 2009, A\&A, 507, 1585

\bibitem[\protect\citeauthoryear{Ritchie et al.}{2010}]{ritchie10} Ritchie, B.W., Clark, J.S., Negueruela, I., \& Langer, N. 2010, A\&A, 520, A48

\bibitem[\protect\citeauthoryear{Rochau et al.}{2010}]{rochau10} Rochau, B., Brandner, W., Stolte, A., et al. 2010, ApJ, 716, L90

\bibitem[\protect\citeauthoryear{Roman-Lopes et al.}{2011}]{rl11} Roman-Lopes, A., Barba, R.H., \& Morrell, N.I. 2011, MNRAS, in press

\bibitem[\protect\citeauthoryear{Sana et al.}{2008}]{sana08} Sana, H., Gosset, E., Naz\'e, Y., et al. 2008, MNRAS, 386, 447

\bibitem[Schaller et al.(1992)]{schaller}Schaller, G., Schaerer, D.,
  Meynet, G. \& Maeder, A. 1992, A\&AS, 96, 269

\bibitem[\protect\citeauthoryear{Schnurr et al.}{2008}]{schnurr08} Schnurr, O., Casoli, J., Chen\'e, A.-N., et al. 2008, MNRAS 389, L38

%\bibitem[\protect\citeauthoryear{Slesnick et al.}{2002}]{sles02} Slesnick, C.L., Hillenbrand, L.A., Massey, P. 2002 , ApJ, 576, 880

\bibitem[\protect\citeauthoryear{Smartt}{2009}]{smartt} Smartt, S.J. 2009, ARA\&A, 47, 63

\bibitem[\protect\citeauthoryear{Smartt et al.}{2009}]{smarttal09} Smartt, S.J., Eldridge, J.J., Crockett, R.M., \& Maund, J.R. 2009, MNRAS, 395, 1409

\bibitem[\protect\citeauthoryear{Smith et al.}{2011}]{smith11} Smith, N., Li, W., Filippenko, A.V., \& Chornock, R. 2011, MNRAS, 412, 1522

\bibitem[\protect\citeauthoryear{Turner}{1981}]{turner81} Turner, D.G. 1981, AJ, 86, 231 

\bibitem[\protect\citeauthoryear{Walborn}{2002}]{walborn02a} Walborn, N.R. 2002, ASPC, 267, 111

\bibitem[\protect\citeauthoryear{Walborn}{2010}]{walborn10} Walborn, N.R. 2010, ASPC, 425, 45

\bibitem[Walborn \& Fitzpatrick(1990)]{wf90} Walborn, N.R. \& Fitzpatrick, E.L. 1990, PASP, 102, 379

\bibitem[\protect\citeauthoryear{Walborn et al.}{2002}]{walborn02} Walborn, N.R., Howarth, I.D., Lennon, D.J., et al. 2002, AJ, 123, 2754

\bibitem[\protect\citeauthoryear{Zavagno et al.}{2005}]{zavagno05} Zavagno, A., Deharveng, L., Brand, J., et al. 2005, IAUS, 227, 346

\bibitem[\protect\citeauthoryear{Zinnecker \& Yorke}{2007}]{zy07} Zinnecker, H., \& Yorke, H.W. 2007, ARA\&A, 45, 481
%
%
% Do not delete next few lines
\end{thebibliography}

\end{document}